\documentclass[aps,twocolumn,showlabels,showrefs,amsmath,amssymb,pre,superscriptaddress, floatfix, colors]{revtex4}

\usepackage{graphicx}
\usepackage{dcolumn}
\usepackage{bm}
\usepackage{amssymb}
\usepackage{hyperref}
\usepackage{multirow}
\usepackage{color}

\newcommand{\1}{\begin{equation}}
\newcommand{\2}{\end{equation}}
\newcommand{\ea}{\begin{eqnarray}} 
\newcommand{\ee}{\end{eqnarray}}
\newcommand{\4}[2]{{\frac{#1}{#2}}}
 
\newcommand{\Sum}[2]{{\sum\limits_{#1}^{#2}}}

\begin{document}
\title{Activity Induced Synchronization: From Mutual Flocking to Chiral {Self-}{Sorting}}

\author{D. Levis}
\thanks{Equal contributions}
\email[corresponding author ]{demian.levis@epfl.ch}
\affiliation{CECAM Centre Europ\'een de Calcul Atomique et Mol\'eculaire, \'Ecole Polytechnique F\'ed\'erale de Lausanne, Batochime, Avenue Forel 2, 1015 Lausanne, Switzerland}
\affiliation{Departament de F\'isica de la Mat\`eria Condensada, Universitat de Barcelona, Mart\'i i Franqu\`es 1, E08028 Barcelona, Spain}
\affiliation{UBICS University of Barcelona Institute of Complex Systems, 
Mart\'i i Franqu\`es 1, E08028 Barcelona, Spain}
\author{I. Pagonabarraga}
\affiliation{CECAM Centre Europ\'een de Calcul Atomique et Mol\'eculaire, \'Ecole Polytechnique F\'ed\'erale de Lausanne, Batochime, Avenue Forel 2, 1015 Lausanne, Switzerland}
\affiliation{Departament de F\'isica de la Mat\`eria Condensada, Universitat de Barcelona, Mart\'i i Franqu\`es 1, E08028 Barcelona, Spain}
\affiliation{UBICS University of Barcelona Institute of Complex Systems, 
Mart\'i i Franqu\`es 1, E08028 Barcelona, Spain}
\author{B. Liebchen}
\thanks{Equal contributions}
\email[corresponding author ]{liebchen@hhu.de}
\affiliation{SUPA, School of Physics and Astronomy, University of Edinburgh, Edinburgh EH9 3FD, United Kingdom}
\affiliation{Institut f\"{u}r Theoretische Physik II: Weiche Materie, Heinrich-Heine-Universit\"{a}t D\"{u}sseldorf, D-40225 D\"{u}sseldorf, Germany}

\begin{abstract}
Synchronization, the temporal coordination of coupled oscillators,
allows fireflies to flash in unison, 
neurons to fire collectively and human crowds to fall in step on the London millenium bridge.
Here, we interpret active (or self-propelled) chiral microswimmers with a distribution of intrinsic frequencies 
as motile oscillators and show that they can synchronize over very large distances, even for local coupling in 2D. This opposes to 
canonical non-active oscillators on static or time-dependent networks,  leading to
synchronized domains only. 
A consequence of this activity-induced synchronization is the emergence of a 
``mutual flocking phase'', where  particles of opposite chirality cooperate to form superimposed flocks moving 
at a relative angle to each other, providing a chiral active matter 
analogue to the celebrated Toner-Tu phase.
The underlying mechanism employs a 
positive feedback loop involving the two-way coupling between oscillators' phase and self-propulsion, 
and could be exploited as a design principle for synthetic active materials and chiral self-sorting techniques. 
\end{abstract}

\maketitle

Populations of motile entities, from bacteria to synthetic microswimmers, can spontaneously self-organize into phases which are
unattainable in passive matter. 
Examples range from the spectacular murmuration of starlings and bacterial swarming \cite{SumpterBook,  Zhang2010, VicsekRev, MarchettiRev}, 
to collective behavior of synthetic  
self-propelled grains \cite{Narayan2007, Deseigne2010}, emulsion droplets \cite{Sanchez2012} or 
assemblies of robots \cite{Rubenstein2014}. Active colloids \cite{BechingerRev}
in particular, spontaneously form living clusters in low density suspensions which, unlike equilibrium clusters, continuously break up and reform
\cite{BechingerRev, Theurkauff2012, Palacci2013, Levis2014, Soto2014, Liebchen2017ph,Ginot2018}.
Remarkably, when active particles align with their neighbors, they can 
self-organize into a polarly ordered phase, featuring long-range order in 2$d$ \cite{Vicsek1995, Toner1995, Toner1998, Toner2005}.
Thus, motile entities, 
like wildebeests or sheep, can order at long-distances and all run in the same direction, whereas immotile entities
like spins in a 2$d$ magnet 
cannot display (true) long-range 
order when locally coupled \cite{Mermin1966,Kosterlitz1973}. 
Likewise, oscillators which are 
localized in space, like metronomes or neurons \cite{Winfree2001,Arenas2008}, can synchronize only  
{\emph{locally} in 2$d$ or 3$d$} when coupled to their neighbors
\cite{Sakaguchi1987, Daido1988, Hong2005}. 

\begin{figure}
\includegraphics[scale=.45]{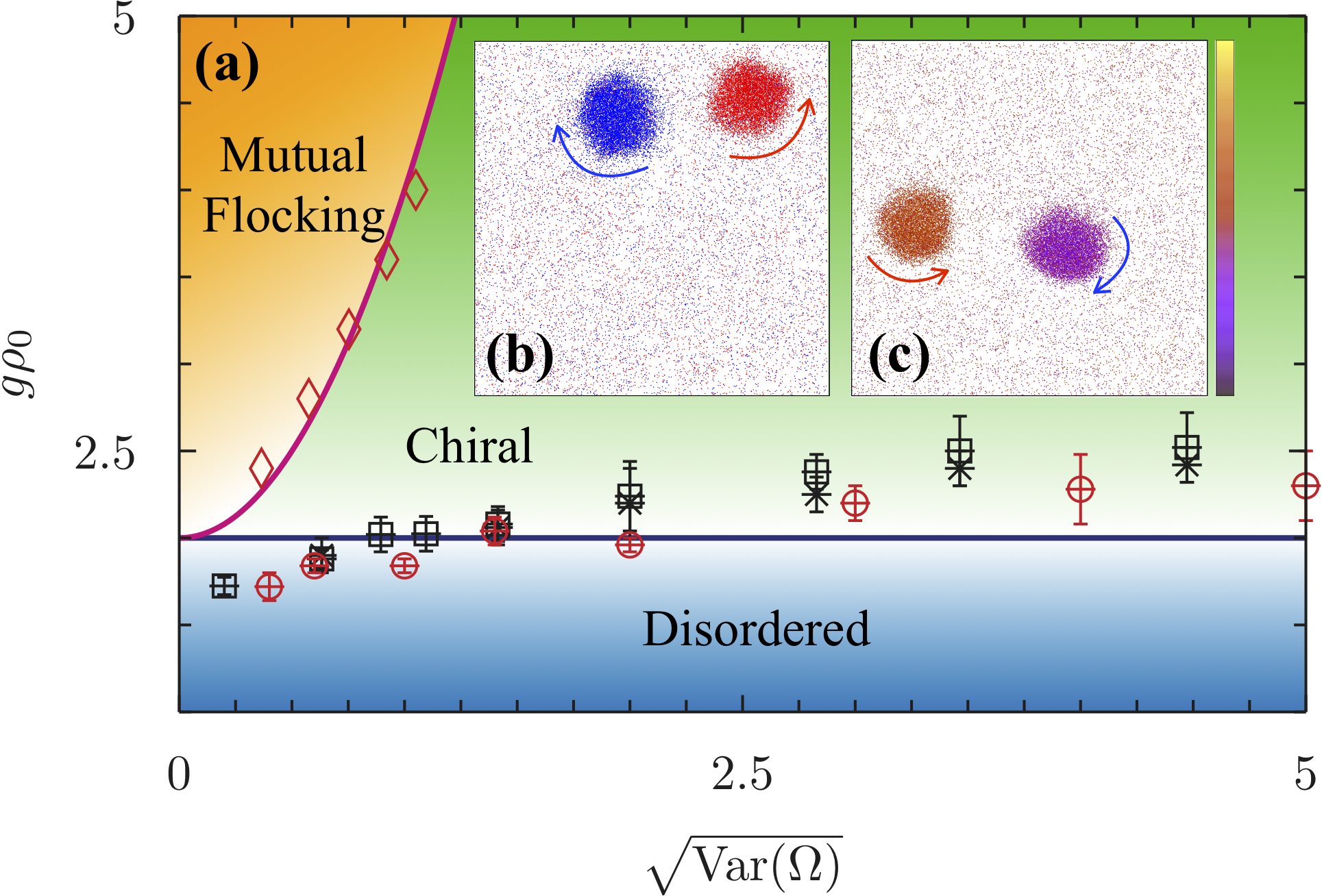}
\caption{
(a)  Phase diagram of active oscillators comprising the mutual flocking phase induced by activity-induced synchronization:
Lines and symbols show analytical predictions and 
simulation for the phase boundaries (red symbols for two species, black ones for a normal frequency distribution) 
using two different densities ($\rho_0=10$, $20$ shown in  black crosses and squares). 
The weak deviation between simulations and predictions probably originates 
from a reduction of the growth rate of unstable modes as $\Omega$ increases
(see SM \cite{SM}).
Insets show snapshots in the chiral phase for two species (b) 
and a Gaussian distribution (c) both for $\mbox{Var}(\Omega)=2; g\rho_0=2.8$; colors show intrinsic frequencies. 
}\label{fig:phd}
\end{figure}
Here, we consider ensembles of chiral active particles with a tendency to swim in circles and interpret them as 'active oscillators', 
to link generic synchronization theories with active matter physics. 
Despite the common occurrence of such 
active oscillators both in nature, e.g.  microorganisms \cite{diLuzio2005, Lauga2006, Shenoy2007, Riedel2005, GompperSperm} or  
cell-components \cite{Sumino2012,Kim2018,Loose2013}, and in the world of synthetic microswimmers
\cite{Kummel2013, Wykes2016,vanTeeffelen2008, Mijalkov2013, Liebchen2016, Lowen2016, Sevilla2016, Liebchen2017}, 
their generic large-scale synchronization behavior remains surprisingly unclear: 
(i) In active matter, previous studies have focused on pattern formation in chiral active particles \cite{Gompper2014, Denk2016,Liebchen2017,Levis2017MF,Liao2018,Lei2018}, but leave the 
synchronization behavior of large ensembles essentially open.
(ii) In the field of synchronization, in turn, 
there is an 
impressive body of knowledge on the large-scale synchronization of oscillators which are localized in space 
\cite{Winfree2001,Arenas2008}, 
or move in a way that is \emph{unaffected} by their phases \cite{Frasca2008, Uriu2014, Grossmann2016, LevisPRX} -- but not for 
the above active-oscillator examples, which show a two-way coupling between phase and displacement through space.

Here, we establish a framework to explore the large-scale synchronization of
active oscillators showing that the two-way coupling, naturally present in active oscillators, creates a novel synchronization behavior 
allowing for synchronization over very large distances. This not only transcends a fundamental limitation of synchronization in locally coupled oscillators
(which can synchronize only {\emph{locally} in 2$d$ or 3$d$} \cite{Sakaguchi1987, Daido1988, Hong2005}), but also 
creates an undiscovered active matter phase, which we call  
the {\it mutual flocking phase}. Here, active oscillators with opposite chirality 
cooperate to move coherently at a relative 
angle, forming non-rotating flocks, akin to the {celebrated} Toner-Tu phase in linear active matter \cite{Toner1995, Toner1998, Toner2005} (see Fig.~\ref{fig:phd}, orange domain). 
This suggests 
that the universality of polar active matter might also apply to active oscillators, 
yet in a remarkably nontrivial way, involving synchronization across non-identical particles (i.e. a distribution of frequencies).

In addition, for comparatively broad frequency 
distributions, we find that 
 {active oscillators} spatially segregate according to their chirality and form 
rotating macroscopic clusters (see Fig.~\ref{fig:phd}, green domain). 
These feature internal synchronization (see snapshots Fig.~\ref{fig:phd} b,c) and grow linearly with system size. 
This {\it chiral phase} constitutes a second example where activity leads to 
synchronization on the macroscale. 
Notice, that the 
chiral self-sorting underlying this phase allows to segregate active partices by chirality, without requiring 
chemical reactions \cite{Gubitz2001}, external flows or environmental chirality \cite{Mijalkov2013}.

To demonstrate the impact of activity on the large-scale synchronization of coupled oscillators, we consider 
$N$ overdamped particles with 2$d$ positions $\boldsymbol{r}_\alpha$ 
and alignment interactions of strength $K$
in a square box of length $L$, which self-propel with a constant speed $v$ along direction
$\boldsymbol{n}_\alpha=(\cos\theta_\alpha,\,\sin\theta_\alpha)$, as described by: 
\begin{eqnarray}
\dot{\boldsymbol{r}}_\alpha&=& v \boldsymbol{n}_\alpha \label{eq:EOM1} \\
\dot{{\theta}}_\alpha &=& \omega_\alpha +\frac{K}{\pi R^2}\sum_{\nu\in\partial_\alpha}\sin(\theta_\nu-\theta_\alpha)+\sqrt{2D_r}{\eta}_\alpha \label{eq:EOM2}
\end{eqnarray}
Here, ${\eta}$ represents Gaussian white noise with unit variance and zero mean. 
The intrinsic 
frequencies $\omega_\alpha$ are randomly drawn from a 
distribution $\Delta(\omega)$.
{Note that, conversely to more complicated models, involving e.g. hydrodynamic interactions, excluded volume effects or sophisticated couplings, the present model 
focuses on the essential ingredients required to demonstrate the 
 generic synchronization scenario shown by active oscillators.
Moreover, direct experimental realizations of the present model are also possible, e.g. using 3D-printed granular particles
on vibrated plates \cite{Deseigne2010,Weber2013,Scholz2018a}. For example, disk-shaped granulates featuring asymmetric legs \cite{Deseigne2010},  both swim in circles and locally align with each other \cite{Deseigne2010,Weber2013}.
Refs.~\cite{Chen2017, Kim2018,Oliver2018} provide other very recent setups containing the key-ingredients of the present model.}
Since we want to understand the synchronization behavior arising from the competition between the distribution of natural frequencies and the coupling, we consider only  
distributions with zero mean. 
{Indeed, here we establish the role played by the frequency dispersion (through its variance $\text{Var}(\omega)$) on the collective behavior of active oscillators, a control parameter that has no equivalent in previous works on chiral active matter.} 
The sum eq. \ref{eq:EOM2} runs over all the neighbors of particle $\alpha$, defined by the cutoff distance {$R$}.
For $v=0$ our model reduces to the noisy Kuramoto model of locally coupled oscillators, which is known to show only local synchronization in 2$d$ \cite{Daido1988,Hong2005}. 
For $\omega_{\alpha}=0$, in turn, the model is equivalent to (a smooth variant of) the Viscek model for which it 
is known that self-propulsion induces long-range order in 2$d$. 
The crucial feature of the present model is that it identifies the oscillators phase with their direction of self-propulsion, giving rise to circular motion at the individual level and 
the emergence of a novel synchronization scenario at the collective level. In the spirit of the Kuramoto model \cite{Acebron2005}  
and simple models of self-propelled particles (like the Vicsek and smooth variants of it \cite{Farrell2012,Peruani2008,Chepizhko2013} or the Active Brownian Particle model \cite{Romanczuk2012, Cates2015, Lino2018}), our model provides a minimal framework to study the generic  behavior emerging in systems of non-identical {active oscillators} or circle swimmers. 

However, most previous studies
on synchronization in dynamic networks have focused on identical oscillators whose phase
does not affect their motion in space \cite{Frasca2008, Uriu2014, Grossmann2016, LevisPRX},  
showing that, the absence of global synchronization (and other scaling properties) for Kuramoto oscillators in 2$d$ is generically preserved, 
even if carried by active particles.
{Note that, despite this recent boost of activity on synchronization of motile oscillators, 
the dispersion of natural frequencies, a  main feature in synchronization problems, has not been considered yet. 
{Therefore, in order to better understand the fundamental synchronization mechanisms involved in active systems, we have } carried extensive simulations of our model in the absence of the back-coupling between the phase of the oscillators and their direction of motion 
 and show that global phase synchronization cannot be achieved (see  SM \cite{SM} for further details). 
Models considering agents whose phase, or a different internal state, affect the way they move in space appeared recently 
\cite{Tanaka2007, Kruk2015, Starnini2016, Keeffe2017}.  
{However, none of the previous studies consider particles self-propelled in the direction of an internal Kuramoto phase (with a distribution of natural frequencies and  local coupling). Then, as expected,  such models cannot account for the generic synchronization scenario of  active oscillators we establish here,  in particular, the possibility to sustain global phase synchronization in 2$d$ (and 3$d$) active matter and provide a route to chiral self-sorting}. 
\begin{figure}
\includegraphics[scale=0.47,angle=0]{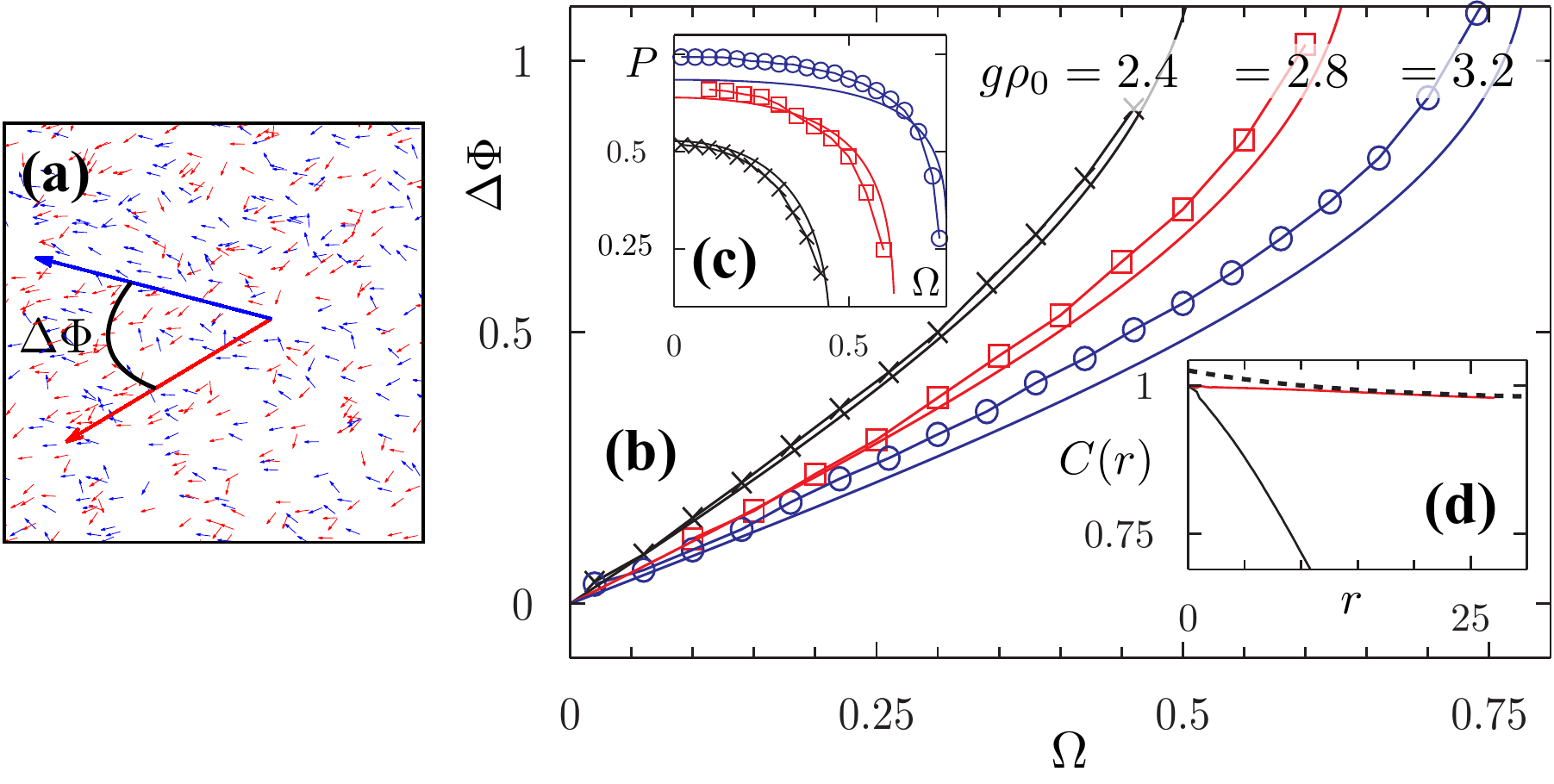}
\caption{Mutual flocking phase.  Detailed view of a late-time configuration for $\Omega=\pm1.5$ at
$g\rho_0=5.6$ (a). Thick arrows show the average polarization of both species, while small arrows show individual particle orientations. 
 Angle $\Delta\Phi(|\Omega|)$ between the two flocks (b) and partial polarization $P(|\Omega|)$ (c) of each species. Lines show eq. \ref{mutflock} and points  simulation results for ${\rm Pe}=2; \rho_0=200$. (d): $C(r)$ (in linear-log scale) of a system of active  (Pe$=2$, in red) and non-motile oscillators (Pe$=0$, in black) for the same set of parameters ($\text{Var}(\Omega)=0.4$ and $g\rho_0=40$). Dotted line shows an exponential decay to the asymptotic value $P^2$.   
}\label{fig:mutualflock}
\end{figure} 
To understand the key control parameters of the present system, we express times and lengths in units of $1/D_r$ and $R$, respectively, 
leading to the dimensionless quantities:  
(i) $g=K/(\pi R^2 D_r)$, the reduced coupling; 
(ii) ${\rm Pe}=v/(D_r R)$, the rotational P\'eclet number; 
(iii) $\Omega_i=\omega_i/D_r$, the reduced rotation frequency, and (iv) $\rho_0=N R^2/L^2$, the average number density.
We denote by $\rho^{(i)}=N^{(i)} R^2/L^2$ the density of particles with natural frequency $\Omega_i$ (species $i$), 
where $N^{(i)}$ counts particles sharing the same natural frequency. 
For simplicity, we exemplify our key results for two species with equal overall density and opposite chirality. 
As we will show, the synchronization behavior is controlled by $\text{Var}(\Omega)$, such that this distribution is 
largely representative to cases of several species and continuous frequency distributions.
\paragraph*{Field Theory}
We first derive a set of field equations describing the collective dynamics of chiral active particles. 
Here, we assume that the system contains $M$ different species with frequencies $\Omega_i$ ($i=1..M$) and describe the dynamics of species-$i$-particles 
using the density field $\rho\equiv \rho^{(i)}({\bf x},t)$
and the polarization density
${\bf w} \equiv {\bf w}^{(i)}({\bf x},t)$,
where ${\bf w}/|{\bf w}|$ is the average self-propulsion direction of particles of species $i$.
The resulting equations read (see SM for details \cite{SM}): 
\begin{eqnarray}
\dot \rho &=& -{\rm Pe} \nabla \cdot {\bf w} \label{rho1}\\
\dot {\bf w}& =& -{\bf w} 
+ \Sum{i=1}{M}\4{g\rho}{2}{\bf w}^{(i)} + \Omega {\bf w}_\perp - \4{{\rm Pe}}{2}\nabla \rho \nonumber \\ 
&+& \4{{\rm Pe}^2}{2 b} \nabla^2 {\bf w} + \4{{\rm Pe}^2 \Omega}{4b}\nabla^2 {\bf w}_\perp \nonumber \\
&-& \4{g^2}{b}\left({\bf w} +\4{\Omega {\bf w}_\perp}{2}\right) \left(\Sum{i=1}{M}{\bf w}^{(i)}\right)^2
+\mathcal{O}\left({\nabla {\bf w}^2}\right) \label{weq}
\end{eqnarray}
Here, $b=2(4+\Omega^2)$, ${{\bf w}_\perp =(-w_y,w_x)}$ and $\mathcal{O}\left(\nabla {\bf w}^2\right)$ represents 
terms which are both nonlinear in ${\bf w}$ and involve gradients, and are not of interest for our following purposes.  
Eq.~(\ref{rho1}) simply reflects that particles on average self-propel in the direction of ${\bf w}$. 
The first term in Eq.~(\ref{weq}) represents a decay of the polarization due to rotational diffusion of the particles, which happens in competition with the 
second term creating alignment among all species; the third term represents a rotation of the average (local) polarization direction 
with a species-specific frequency while the 
forth term expresses a statistical tendency for self-propulsion away from high-density regions; remaining terms 
'smear out' regions of high and low polarization and lead to saturation.
For weak  interactions, the system is in a disordered uniform phase, 
given by $(\rho,{\bf w})=(\rho_0,{\bf 0})$, which is a solution of Eqs.~(\ref{rho1}, \ref{weq}) and represented by the 
blue domain in the phase diagram Fig.~\ref{fig:phd}.

\paragraph*{Mutual Flocking Phase}
To understand the onset of synchronization we
perform a linear stability analysis for a bimodal frequency distribution $\Omega_1=-\Omega_2{=\Omega}$
of the uniform disordered phase ($\rho^{(1),(2)}=\rho_0/2$, ${\bf w}_1={\bf w}_2=0$) \cite{SM}. 
It is instructive to first focus on the zero wavenumber limit ($q=0$).
Here, we find that the uniform phase looses stability if $g\rho_0>2(1+\Omega^2)$, 
suggesting the existence of a nontrivial ordered uniform phase.
We indeed find an  
exact solution of our field equations (\ref{rho1},\ref{weq}) representing such a phase: 
 a state
of uniform density $\rho^{(1)}=\rho^{(2)}=\rho_0/2$, but with a finite polarization, spontaneously breaking the symmetry of the disordered phase. 
In this state,  circle swimmers of both species cooperate and form two individual and superimposed 
flocks moving linearly and at a relative angle $\Delta\Phi $ to each other. We thus call it \emph{mutual flocking phase}.
Defining
the overall polarization $P=|{\bf w}_1+{\bf w}_2|/\rho_0$, where $|{\bf w}_1|=|{\bf w}_2|=\rho_0 P/\sqrt{2(1+\cos\Delta\Phi )}$, the exact expressions 
representing the mutual flocking phase read
\begin{eqnarray}
P &=& \frac{\sqrt{2}}{{g \rho_0}} \sqrt{\sqrt{(g \rho_0)^2+6 \Omega ^2 (g \rho_0 -6)}+g \rho_0 +2 \Omega ^2-4}  \label{mutflock} \\
\Delta\Phi &=& -i \ln \left[\frac{2\Omega \left(6-\frac{g \rho_0}{2}\right)+4 i \sqrt{\frac{(g \rho_0)^2}{4}+3 \Omega^2 \left(\frac{g \rho_0}{2}-3\right)}}{g \rho _0 \left(\Omega+2 i\right)}\right] \nonumber 
\end{eqnarray}
At low frequency $\Omega<1/\sqrt{2}$, this solution exists (is real and positive) if $g\rho_0>2(1+\Omega^2)$, i.e. precisely when the field equations show a linear instability at $q=0$.

 Brownian dynamics simulations at high coupling and high density confirm the existence of the mutual flocking phase. 
As shown in Fig.~\ref{fig:phd} (a), we find the mutual flocking phase essentially in the whole parameter regime where it exists according to our field theory. 
Fig. \ref{fig:mutualflock}  shows in turn a close quantitative agreement between theoretical predictions (Eqs. \ref{mutflock}) and 
our simulations, both for the angle between the two flocks and for the overall polarization 
 (see Movie 1 \cite{SM}). 
(In our simulations, we measure the partial and overall polarization, $\tilde{P}=|\bold{P}_{1,2}|$, 
where $\bold{P}_{i}=\frac{2}{N}\sum_{\alpha}\bold{n}_{\alpha}\delta_{\omega_{\alpha},\omega_{i}}$, and $P=\frac{1}{N}|\sum_{\alpha}\bold{n}_{\alpha}|$, respectively.)  In Fig. \ref{fig:mutualflock} (d) we show the orientational self-correlation function  $C(r)=\langle \boldsymbol{n}_i\cdot \boldsymbol{n}_j\rangle$, strongly suggesting the emergence of global synchronization for active oscillators 
in contrast to immobile oscillators ($v=0$ in eq. \ref{eq:EOM1}). 

\paragraph*{Activity-induced synchronization}
To generally understand when the disordered phase looses stability, we now explore its linear stability at finite wavenumber $q\neq 0$. This is 
equivalent to accounting for the impact of motility  on the onset of synchronization, as the coefficients of all gradient terms in \eqref{rho1}, \eqref{weq} are non-vanishing only if ${\rm Pe}\neq0$. 
While passive oscillators need a stronger coupling to synchronize as the 
frequency dispersion increases \cite{Acebron2005, Arenas2008}, our 
  linear stability calculation shows that 
self-propulsion generally induces an instability for any $g\rho_0>2$ \cite{SM}, independently of $\text{Var}(\Omega)$.
That is, self-propulsion induces phase ordering even in parameter regimes where a 
corresponding non-motile system (${\rm Pe}=0$) would simply show asynchronous 
oscillations of the individual particles. 
As the present instability emerges at finite $q$, i.e. for localized perturbations, we do not expect a uniform phase, but rather the emergence of
localized synchronous structures. Our particle based simulations confirm this. 
Above the critical value $g\rho_0 \gtrsim 2$ [Fig.~\ref{fig:phd} (a)], we find large rotating clusters of opposite chirality featuring internal phase-synchronization, 
as illustrated by the snapshots Fig.~\ref{fig:phd} (b-c) (and Movie 2 \cite{SM}).
From   
$C(r)$ 
we extract a characteristic length $\xi$, which, as shown in Fig. \ref{fig:seg} (a), 
grows linearly with $\sqrt{N}$ (at fixed $\rho_0$), meaning that clusters are macroscopic in this regime and tend to phase separate. 
Thus, the chiral phase represents a second example where self-propulsion induces (long-range) phase-synchronization among each species.
As shown in Fig. \ref{fig:seg} (b), above a critical coupling strength, $\tilde{P}$ starts to increase, allowing us to locate the phase boundaries Fig. 
\ref{fig:phd} (a) (Particle segregation, quantified by the probability $\Psi$ to find an excess of particles of one chirality, 
offers an alternative route to identify the emergence of the chiral phase, see \cite{SM}).
\begin{figure}
\includegraphics[scale=0.45]{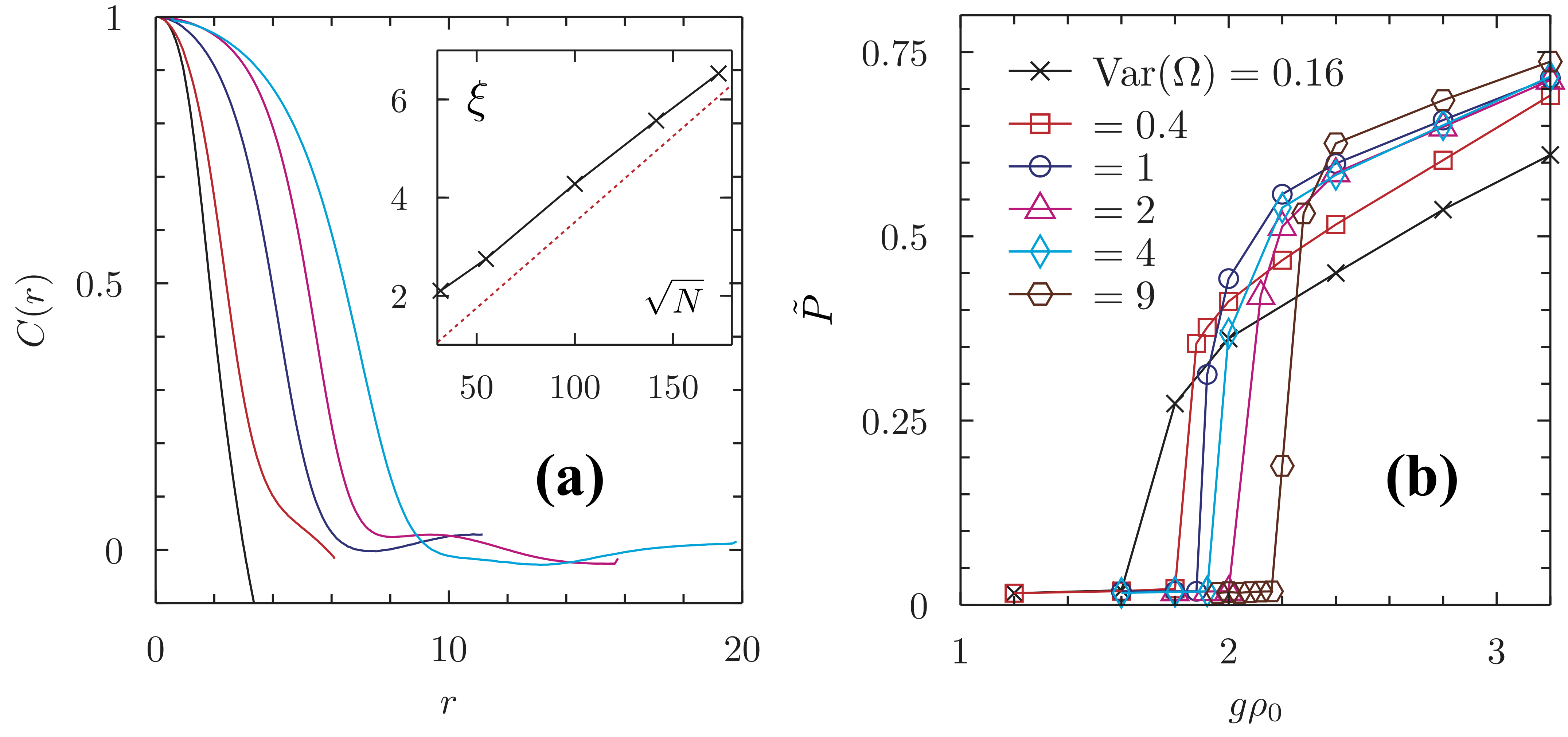}
\caption{ Chiral phase:
(a): Spatial correlations for $g\rho_0=2.8$, $\mbox{Var}(\Omega)=2$  and several system sizes ($N=10^3$, ..., $3.10^4$, from left to right).  The inset shows the corresponding correlation length $\xi$ as a function of $\sqrt{N}$.  We show for comparison a linear growth $\propto \sqrt{N}$ in dotted line.  
(b): Partial polarization $\tilde{P}$  as a function of   $g\rho_0$ for several frequency dispersions.  
}\label{fig:seg}
\end{figure}

\paragraph*{Continous frequency distributions}
To demonstrate the generality of our results, 
we now simulate active oscillators with Gaussian distribution of natural frequencies (with zero mean and variance $\text{Var}(\Omega)$). 
As for two species, we predict the stability threshold of the disordered phase at $g\rho_0>2$ for 3 and 4 species using linear stability analysis 
(see SM \cite{SM}) and confirm with simulations that the stability threshold of the disordered state does not change as compared to the 2-species case, up to numerical accuracy (Fig. \ref{fig:phd}).
At $g\rho_0 \gtrsim 2$, we find the same two phases as for two species: a mutual flocking phase, further illustrated in the SM \cite{SM}, 
and a phase comprising synchronized macroclusters scaling linearly with system size (see Fig. \ref{fig:phd} (c) and movie 3 \cite{SM}). 
Now, the macroclusters involve a continuos range of frequencies and feature frequency synchronization across species.

\paragraph*{Physical Mechanism}
What is the physical mechanism allowing motility to qualitatively change the synchronization of oscillators? 
The absence of global synchronization is well-known 
for static structures of dimension $d<d_c=4$ (where $d_c$ denotes its lower critical dimension) 
\cite{Sakaguchi1987, Daido1988, Hong2005}, and 
particle motion itself does not affect such scenario 
if the oscillators motion is independent of their phase \cite{LevisPRX, SM}.
Hence, compared to oscillators on dynamic networks, the crucial new ingredient here is that the phase of an oscillator determines its direction of motion. 
It follows that two circle swimmers sharing the same phase are aligned and move together, enhancing their interaction time, which in turn enhances their alignment and fosters synchronization. 
This can be viewed as a positive feedback between alignment (local synchronization) and interaction-time, mutually supporting each other, which is 
absent for oscillators on dynamic networks but plays a fundamental role for synchronization of active oscillators. 
 The impact of such feedback
on the emergence of order is even more dramatic in the present case than in linear active matter. 
Unlike the static XY-model for immotile spins,  featuring quasi-long-range order in $d=d_c=2$, 
the Vicsek model (and its continuous variants) shows long-range order in the form of a flocking (Toner-Tu) phase. 
Although for phase-oscillators (Kuramoto model) 
 $d_c=4$, global phase synchronization occurs in $d=2$ for active oscillators, well below the critical dimension of its passive counterpart. 

\paragraph*{Conclusions} 
Chiral active particles, here interpreted as 2D active oscillators, can synchronize over very long distances, even for a purely local coupling. 
This contrasts the synchronization behavior of the  
huge class of non-active oscillators in 2$d$ or 3$d$ structures \cite{Arenas2008}, which can only synchronize within domains.
A consequence of global synchronization in active oscillators, 
is the emergence of the mutual flocking phase as a new active matter phase, akin to the celebrated Toner-Tu phase in linear active matter.
Our results  
transcend a knowledge boundary at the interface of active matter physics and synchronization and could be useful e.g. to sort 
active enantiomers.

\textbf{Acknowledgements} The authors thank H. L\"owen and D. Marenduzzo for fruitful discussions. DL and BL acknowledge received funding from a Marie Curie Intra European Fellowship (G. A. No. 654908 and G. A. No. 657517) within Horizon 2020. IP acknowledges MINECO and DURSI for financial
support under projects FIS2015-67837- P and 2017SGR-884,
respectively.


\bibliographystyle{unsrt}
\bibliography{SyncCAP02}

\begin{thebibliography}{10}

\bibitem{SumpterBook}
D~J~T Sumpter.
\newblock {\em Collective animal behavior}.
\newblock Princeton University Press, 2010.

\bibitem{Zhang2010}
H-P Zhang, A~Beer, E-L Florin, and H~L Swinney.
\newblock Collective motion and density fluctuations in bacterial colonies.
\newblock {\em Proc. Nat. Ac. Sci. USA}, 107(31):13626--13630, 2010.

\bibitem{VicsekRev}
T~Vicsek and A~Zafeiris.
\newblock Collective motion.
\newblock {\em Phys. Rep.}, 517(3):71, 2012.

\bibitem{MarchettiRev}
M~C Marchetti, J-F Joanny, S~Ramaswamy, T~B Liverpool, J~Prost, M~Rao, and R~A
  Simha.
\newblock Hydrodynamics of soft active matter.
\newblock {\em Rev. Mod. Phys.}, 85(3):1143, 2013.

\bibitem{Narayan2007}
V~Narayan, S~Ramaswamy, and N~Menon.
\newblock Long-lived giant number fluctuations in a swarming granular nematic.
\newblock {\em Science}, 317(5834):105--108, 2007.

\bibitem{Deseigne2010}
J~Deseigne, O~Dauchot, and H~Chat{\'e}.
\newblock Collective motion of vibrated polar disks.
\newblock {\em Phys. Rev. Lett.}, 105(9):098001, 2010.

\bibitem{Sanchez2012}
T~Sanchez, D~TN Chen, S~J DeCamp, M~Heymann, and Z~Dogic.
\newblock Spontaneous motion in hierarchically assembled active matter.
\newblock {\em Nature}, 491(7424):431--434, 2012.

\bibitem{Rubenstein2014}
M~Rubenstein, A~Cornejo, and R~Nagpal.
\newblock Programmable self-assembly in a thousand-robot swarm.
\newblock {\em Science}, 345(6198):795--799, 2014.

\bibitem{BechingerRev}
C~Bechinger, R~Di~Leonardo, H~L{\"o}wen, C~Reichhardt, G~Volpe, and G~Volpe.
\newblock Active particles in complex and crowded environments.
\newblock {\em Rev. Mod. Phys.}, 88(4):045006, 2016.

\bibitem{Theurkauff2012}
I~Theurkauff, C~Cottin-Bizonne, J~Palacci, C~Ybert, and L~Bocquet.
\newblock Dynamic clustering in active colloidal suspensions with chemical
  signaling.
\newblock {\em Phys. Rev. Lett.}, 108(26):268303, 2012.

\bibitem{Palacci2013}
J~Palacci, S~Sacanna, A~P Steinberg, D~J Pine, and P~M Chaikin.
\newblock Living crystals of light-activated colloidal surfers.
\newblock {\em Science}, 339(6122):936, 2013.

\bibitem{Levis2014}
D~Levis and L~Berthier.
\newblock Clustering and heterogeneous dynamics in a kinetic monte carlo model
  of self-propelled hard disks.
\newblock {\em Phys. Rev. E}, 89(6):062301, 2014.

\bibitem{Soto2014}
R~Soto and R~Golestanian.
\newblock Run-and-tumble dynamics in a crowded environment: Persistent
  exclusion process for swimmers.
\newblock {\em Phys. Rev. E}, 89(1):012706, 2014.

\bibitem{Liebchen2017ph}
B~Liebchen, D~Marenduzzo, and M~E Cates.
\newblock Phoretic interactions generically induce dynamic clusters and wave
  patterns in active colloids.
\newblock {\em Phys. Rev. Lett.}, 118:268001, 2017.

\bibitem{Ginot2018}
F~Ginot, I~Theurkauff, F~Detcheverry, C~Ybert, and C~Cottin-Bizonne.
\newblock Aggregation-fragmentation and individual dynamics of active clusters.
\newblock {\em Nat. Commun.}, 9(1):696, 2018.

\bibitem{Vicsek1995}
T~Vicsek, A~Czir{\'o}k, E~Ben-Jacob, I~Cohen, and O~Shochet.
\newblock Novel type of phase transition in a system of self-driven particles.
\newblock {\em Phys. Rev. Lett.}, 75(6):1226, 1995.

\bibitem{Toner1995}
J~Toner and Y~Tu.
\newblock Long-range order in a two-dimensional dynamical xy model: how birds
  fly together.
\newblock {\em Phys. Rev. Lett.}, 75(23):4326, 1995.

\bibitem{Toner1998}
J~Toner and Y~Tu.
\newblock Flocks, herds, and schools: A quantitative theory of flocking.
\newblock {\em Phys. Rev. E}, 58(4):4828, 1998.

\bibitem{Toner2005}
J~Toner, Y~Tu, and S~Ramaswamy.
\newblock Hydrodynamics and phases of flocks.
\newblock {\em Ann. Phys.}, 318(1):170, 2005.

\bibitem{Mermin1966}
N~D Mermin and H~Wagner.
\newblock Absence of ferromagnetism or antiferromagnetism in one-or
  two-dimensional isotropic heisenberg models.
\newblock {\em Phys. Rev. Lett.}, 17(22):1133, 1966.

\bibitem{Kosterlitz1973}
J~M Kosterlitz and D~J Thouless.
\newblock Ordering, metastability and phase transitions in two-dimensional
  systems.
\newblock {\em J. Phys. C}, 6(7):1181, 1973.

\bibitem{Winfree2001}
A~T Winfree.
\newblock {\em The geometry of biological time}, volume~12.
\newblock Springer Science \& Business Media, 2001.

\bibitem{Arenas2008}
A~Arenas, A~D\'{\i}az-Guilera, J~Kurths, Y~Moreno, and C~Zhou.
\newblock {Synchronization in complex networks}.
\newblock {\em Phys. Rep.}, 469(3):93, 2008.

\bibitem{Sakaguchi1987}
H~Sakaguchi, S~Shinomoto, and Y~Kuramoto.
\newblock Local and grobal self-entrainments in oscillator lattices.
\newblock {\em Prog. Theor. Phys.}, 77(5):1005--1010, 1987.

\bibitem{Daido1988}
H~Daido.
\newblock Lower critical dimension for populations of oscillators with randomly
  distributed frequencies: a renormalization-group analysis.
\newblock {\em Phys. Rev. Lett.}, 61(2):231, 1988.

\bibitem{Hong2005}
H~Hong, H~Park, and M~Y Choi.
\newblock Collective synchronization in spatially extended systems of coupled
  oscillators with random frequencies.
\newblock {\em Phys. Rev. E}, 72(3):036217, 2005.

\bibitem{SM}
{\em See Supplementary Material at doi:...}

\bibitem{diLuzio2005}
W~R DiLuzio, L~Turner, M~Mayer, P~Garstecki, Douglas~B Weibel, H~C Berg, and
  G~M Whitesides.
\newblock Escherichia coli swim on the right-hand side.
\newblock {\em Nature}, 435(7046):1271, 2005.

\bibitem{Lauga2006}
E~Lauga, W~R DiLuzio, G~M Whitesides, and H~A Stone.
\newblock Swimming in circles: motion of bacteria near solid boundaries.
\newblock {\em Biophys. J.}, 90(2):400, 2006.

\bibitem{Shenoy2007}
V~B Shenoy, D~T Tambe, A~Prasad, and J~A Theriot.
\newblock A kinematic description of the trajectories of listeria monocytogenes
  propelled by actin comet tails.
\newblock {\em Proc. Natl. Acad. Sci.}, 104(20):8229, 2007.

\bibitem{Riedel2005}
I~H Riedel, K~Kruse, and J~Howard.
\newblock A self-organized vortex array of hydrodynamically entrained sperm
  cells.
\newblock {\em Science}, 309(5732):300, 2005.

\bibitem{GompperSperm}
Y~Yang, J~Elgeti, and G~Gompper.
\newblock Cooperation of sperm in two dimensions: synchronization, attraction,
  and aggregation through hydrodynamic interactions.
\newblock {\em Phys. Rev. E}, 78(6):061903, 2008.

\bibitem{Sumino2012}
Y~Sumino, K~H Nagai, Y~Shitaka, D~Tanaka, K~Yoshikawa, H~Chat{\'e}, and K~Oiwa.
\newblock Large-scale vortex lattice emerging from collectively moving
  microtubules.
\newblock {\em Nature}, 483(7390):448--452, 2012.

\bibitem{Kim2018}
K~Kim, N~Yoshinaga, S~Bhattacharyya, H~Nakazawa, M~Umetsu, and W~Teizer.
\newblock Large-scale chirality in an active layer of microtubules and kinesin
  motor proteins.
\newblock {\em Soft Matter}, 14:3221, 2018.

\bibitem{Loose2013}
M~Loose and T~J Mitchison.
\newblock The bacterial cell division proteins ftsa and ftsz self-organize into
  dynamic cytoskeletal patterns.
\newblock {\em Nat. Cell Biol.}, 16(1):38, 2014.

\bibitem{Kummel2013}
F~K{\"u}mmel, B~ten Hagen, R~Wittkowski, I~Buttinoni, R~Eichhorn, G~Volpe,
  H~L{\"o}wen, and C~Bechinger.
\newblock Circular motion of asymmetric self-propelling particles.
\newblock {\em Phys. Rev. Lett.}, 110(19):198302, 2013.

\bibitem{Wykes2016}
M~S~D Wykes, J~Palacci, T~Adachi, L~Ristroph, X~Zhong, M~D Ward, J~Zhang, and
  M~J Shelley.
\newblock Dynamic self-assembly of microscale rotors and swimmers.
\newblock {\em Soft Matter}, 12(20):4584, 2016.

\bibitem{vanTeeffelen2008}
S~van Teeffelen and H~L{\"o}wen.
\newblock Dynamics of a brownian circle swimmer.
\newblock {\em Phys. Rev. E}, 78(2):020101, 2008.

\bibitem{Mijalkov2013}
M~Mijalkov and G~Volpe.
\newblock Sorting of chiral microswimmers.
\newblock {\em Soft Matter}, 9(28):6376--6381, 2013.

\bibitem{Liebchen2016}
B~Liebchen, M~E Cates, and D~Marenduzzo.
\newblock Pattern formation in chemically interacting active rotors with
  self-propulsion.
\newblock {\em Soft Matter}, 12(35):7259--7264, 2016.

\bibitem{Lowen2016}
H~L{\"o}wen.
\newblock Chirality in microswimmer motion: From circle swimmers to active
  turbulence.
\newblock {\em Eur. Phys. J. Spec. Top.}, 225(11):2319, 2016.

\bibitem{Sevilla2016}
F~J Sevilla.
\newblock Diffusion of active chiral particles.
\newblock {\em Phys. Rev. E}, 94(6):062120, 2016.

\bibitem{Liebchen2017}
B~Liebchen and D~Levis.
\newblock Collective behavior of chiral active matter: Pattern formation and
  enhanced flocking.
\newblock {\em Phys. Rev. Lett.}, 119(5):058002, 2017.

\bibitem{Gompper2014}
Y~Yang, F~Qiu, and G~Gompper.
\newblock Self-organized vortices of circling self-propelled particles and
  curved active flagella.
\newblock {\em Physical Review E}, 89(1):012720, 2014.

\bibitem{Denk2016}
J~Denk, L~Huber, E~Reithmann, and E~Frey.
\newblock Active curved polymers form vortex patterns on membranes.
\newblock {\em Phys. Rev. Lett.}, 116(17):178301, 2016.

\bibitem{Levis2017MF}
D~Levis and B~Liebchen.
\newblock Micro-flock patterns and macro-clusters in chiral active brownian
  disks.
\newblock {\em J. Phys.: Condens. Matter}, 30:084001, 2018.

\bibitem{Liao2018}
G-J Liao and S~HL Klapp.
\newblock Clustering and phase separation of circle swimmers dispersed in a
  monolayer.
\newblock {\em Soft matter}, 14(38):7873, 2018.

\bibitem{Lei2018}
Q-L Lei, M~P Ciamarra, and R~Ni.
\newblock Non-equilibrium strong hyperuniform fluids of athermal active circle
  swimmers with giant local fluctuations.
\newblock {\em arXiv preprint arXiv:1802.03682}, 2018.

\bibitem{Frasca2008}
M~Frasca, A~Buscarino, A~Rizzo, L~Fortuna, and S~Boccaletti.
\newblock Synchronization of moving chaotic agents.
\newblock {\em Phys. Rev. Lett.}, 100(4):044102, 2008.

\bibitem{Uriu2014}
K~Uriu and L~G Morelli.
\newblock Collective cell movement promotes synchronization of coupled genetic
  oscillators.
\newblock {\em Biophys. J.}, 107(2):514, 2014.

\bibitem{Grossmann2016}
R~Gro{\ss}mann, F~Peruani, and M~B{\"a}r.
\newblock Superdiffusion, large-scale synchronization, and topological defects.
\newblock {\em Phys. Rev. E}, 93(4):040102, 2016.

\bibitem{LevisPRX}
D~Levis, I~Pagonabarraga, and A~Diaz-Guilera.
\newblock Synchronization in dynamical networks of locally coupled
  self-propelled oscillators.
\newblock {\em Phys. Rev. X}, 7(1):011028, 2017.

\bibitem{Gubitz2001}
G~G{\"u}bitz and M~G Schmid.
\newblock Chiral separation by chromatographic and electromigration techniques.
\newblock {\em Biopharm. Drug. Dispos.}, 22(7-8):291--336, 2001.

\bibitem{Weber2013}
C~A Weber, T~Hanke, J~Deseigne, S~L{\'e}onard, O~Dauchot, E~Frey, and
  H~Chat{\'e}.
\newblock Long-range ordering of vibrated polar disks.
\newblock {\em Phys. Rev. Lett.}, 110(20):208001, 2013.

\bibitem{Scholz2018a}
C~Scholz, M~Engel, and T~P{\"o}schel.
\newblock Rotating robots move collectively and self-organize.
\newblock {\em Nat. Comm.}, 9(1):931, 2018.

\bibitem{Chen2017}
C~Chen, S~Liu, X-q Shi, H~Chat{\'e}, and Y~Wu.
\newblock Weak synchronization and large-scale collective oscillation in dense
  bacterial suspensions.
\newblock {\em Nature}, 542(7640):210, 2017.

\bibitem{Oliver2018}
A~Barroso L Dewenter M~Woerdemann N~Oliver, C~Alpmann and C~Denz.
\newblock Synchronization in pairs of rotating active biomotors.
\newblock {\em Soft Matter, Accepted Manuscript}.

\bibitem{Acebron2005}
J~A Acebr{\'o}n, L~L Bonilla, C~J P{\'e}rez-Vicente, F~Ritort, and R~Spigler.
\newblock The kuramoto model: A simple paradigm for synchronization phenomena.
\newblock {\em Rev. Mod. Phys.}, 77:137, 2005.

\bibitem{Farrell2012}
F~D~C Farrell, M~C Marchetti, D~Marenduzzo, and J~Tailleur.
\newblock Pattern formation in self-propelled particles with density-dependent
  motility.
\newblock {\em Phys. Rev. Lett.}, 108(24):248101, 2012.

\bibitem{Peruani2008}
F~Peruani, A~Deutsch, and M~B{\"a}r.
\newblock A mean-field theory for self-propelled particles interacting by
  velocity alignment mechanisms.
\newblock {\em Eur. Phys. J. Spec. Top.}, 157(1):111, 2008.

\bibitem{Chepizhko2013}
O~Chepizhko and F~Peruani.
\newblock Diffusion, subdiffusion, and trapping of active particles in
  heterogeneous media.
\newblock {\em Phys. Rev. Lett.}, 111(16):160604, 2013.

\bibitem{Romanczuk2012}
P~Romanczuk, M~B{\"a}r, W~Ebeling, B~Lindner, and L~Schimansky-Geier.
\newblock Active brownian particles.
\newblock {\em Eur. Phys. J. Spec. Top.}, 202(1):1, 2012.

\bibitem{Cates2015}
M~E Cates and J~Tailleur.
\newblock Motility-induced phase separation.
\newblock {\em Annu. Rev. Condens. Matter Phys.}, 6(1):219, 2015.

\bibitem{Lino2018}
P~Digregorio, D~Levis, A~Suma, G~Gonnella, L~Cugliandolo, and I~Pagonabarraga.
\newblock Full phase diagram of self-propelled hard-disks: from melting to
  motility induced phase separation.
\newblock {\em Phys. Rev. Lett.}, 121:098003, 2018.

\bibitem{Tanaka2007}
D~Tanaka.
\newblock General chemotactic model of oscillators.
\newblock {\em Phys. Rev. Lett.}, 99(13):134103, 2007.

\bibitem{Kruk2015}
N~Kruk, Y~Maistrenko, N~Wenzel, and H~Koeppl.
\newblock Self-propelled chimeras.
\newblock {\em arXiv preprint arXiv:1511.04738}, 2015.

\bibitem{Starnini2016}
M~Starnini, M~Frasca, and A~Baronchelli.
\newblock Emergence of metapopulations and echo chambers in mobile agents.
\newblock {\em Sci. Rep.}, 6:31834, 2016.

\bibitem{Keeffe2017}
K~P O~'~Keeffe, H~Hong, and S~H Strogatz.
\newblock Oscillators that sync and swarm.
\newblock {\em Nature Comm.}, 8(1):1504, 2017.

\end{thebibliography}
\end{document}